\documentstyle[12pt]{article}
\newcommand{\be}{\begin{equation}}
\newcommand{\ee}{\end{equation}}
\newcommand{\beq}{\begin{eqnarray}}
\newcommand{\eeq}{\end{eqnarray}}

\begin{document}
\vskip .20in
\date{September 2003\\
Brown-HET-1377}
\title
{\bf{Possibilities of a QED-Based Vacuum Energy} }
\vskip .15in
\author{H. M. FRIED\\
{\it Department of Physics} \\
{\it Brown University} \\
{\it Providence R.I. 02912 USA}}  
\vskip .20in
\maketitle
\abstract{
	A QED-based bootstrap mechanism, appearing at sufficiently 
	small space-time scales, 
is suggested as an explanation for the 
``dark" vacuum 
energy that may be able to accelerate
the universe.  Very-small-scale virtual vacuum currents 
are assumed to generate small-scale
electromagnetic fields, corresponding to the appearance 
of an effective 4-potential $A_{\mu }^{ext}
(x)$, which is itself equal to the vev of the operator $A_{\mu}(x)$ 
in the presence of that
$A_{\mu}^{ext}(x)$. The latter condition generates a bootstrap-like 
equation for $A_{\mu}^{ext}(x)$, which has
an approximate, tachyonic-like solution corresponding to 
propagation outside the light cone,
and damping inside; this solution is given in terms of a 
mass parameter $\, M \,$ that turns out to
be on the order of 10$^8$ GeV if all three lepton bubbles are included. 
Virtual quark bubbles
are not expected to be important,  except for the possibility of 
providing the initial
``spark" which sets off the QED bootstrap.

	A multiplicative 4-vector $v_{\mu}$, whose magnitude is
	 determined by a comparison with the 
average mass density needed to produce the observed acceleration 
is introduced, and may
characterize the distance $\,d\,$ over which the fields so produced 
may be expected to be
coherent; the present analysis suggests that d can lie anywhere 
in the range from 10$^{-5}$ cm
(corresponding to a ``spontaneous QED-vacuum phase change") to 10$^{-13}$ 
cm (representing a
``polarization of the QED vacuum" by quark-antiquark pairs of the
 QCD vacuum). Treated as a
constant, or ``averaged" 4-vector in these estimates, $v_{\mu}$ is understood 
to have a
non-perturbative QED or QCD origin. 

	Near the light-cone, such vacuum electric fields become large, 
	suggesting the possibility 
of copious lepton-pair production, resulting in a ``plasma" which tends 
to diminish the
fields.  This suggests a non-perturbative change in the definition of 
the QED vacuum, which
can act to provide a convergence factor for all QED perturbative 
calculations; in essence,
the parameter $M $ defines the scale at which all high-momentum 
Feynman graphs converge.  Four
questions are proposed which carry this non-perturbative, 
vacuum-field mechanism towards
unexpected consequences, involving the possible production 
of tachyonic lepton-pairs that
contribute to ``dark" or ``missing" matter, while acting as a 
source of galactic gamma-ray
bursts and ultra-high-energy cosmic rays.}

{\bf\section{Introduction}}

	Recent experimental results \cite{one} suggesting the outward
	 acceleration of the universe have 
concentrated attention on the vacuum as a possible source of 
self-energy, contributing a
pressure sufficient to produce the measured acceleration.  
In this scenario, Lorentz
invariance (LI) cannot be rigorously true; but since our immediate,
 measured world is almost
perfectly flat, this lack of LI might be expected to play a minor role. 
 A major role could
be played by the mechanism suggested here, in which a lack of 
translational invariance (TI)
appears when the virtual, QED vacuum is examined at ultra-short distances.  
Those distances
range from the inverse of $10^8  GeV$ down to the inverse of, roughly, 
the Planck mass; the
smaller of the two appears if only the simplest, electron vacuum 
bubble is retained, while
the larger is obtained when the muon and tau vacuum bubbles are included. 

	A constant four-vector $v_{\mu}$ is introduced to characterize 
	the distance over which new, ``effectively external" 
	electromagnetic fields are produced and/or 
may be expected to remain
coherent at such very small scales. These fields are generated by 
electrical currents
induced in the vacuum (in the bootstrap-presence of the very 
external fields they produce).
The order-of-magnitude of $v_{\mu}$ is determined by matching a derived
 expression for this
vacuum-induced energy density to the mass density $ U \sim 10^{-29}
 gm/cm^3 $ 
needed to fit
observation;\cite{two} in this computation, the Planck length is used 
as a natural cut-off,
corresponding to the smallest, physically-sensible distances allowed.

	One prejudice of the author should be stated at the outset, 
	which forms part of the 
motivation for the present remarks. The by-now conventional approach 
to vacuum energy is
that the latter represents, in some fashion, zero-point-energies of 
relevant quantum
fields.  Aside from being horribly divergent, and so requiring 
massive renormalizations, it
is not clear that zero-point terms even belong in any field theory, 
for they can be
well-understood as the remnants of improper positioning of products 
of operators at the same
space-time point.  One cannot proceed from classical to quantum 
forms without facing this
question, which has long ago been given a Lorentz-invariant answer 
in terms of ``normal
ordering". When Lorentz invariance is subsumed into a more general 
relativistic invariance,
in which all energies couple to the metric, there is still no 
requirement of including those
zero-point terms which should have been excluded at the very 
beginning; rather, zero-point
terms have been pressed into service as the simplest way of 
attempting to understand vacuum
energy. Because zero-point energies have been used to give a 
qualitative explanation of the
Casimir effect and the Lamb shift does not mean that they are 
the only, or the correct,
explanations for those processes.\cite{three}  It is here suggested 
that there is another source of
vacuum energy, one which violates no principles of quantum 
field theory, that can provide a
simple and at least qualitatively reasonable QED mechanism 
for the acceleration of the
universe, as well as other, unexpected consequences.

{\bf\section{Formulation}}

	Our conventional definition of ``vacuum" is the 
	absence of all ``on-shell" phenomena.  
Further, there can be no doubt of the reality of "off-shell" 
vacuum fluctuations,
demonstrated long ago by the -27 megacycles/sec contribution 
to the $^{2}S_{1/2} - ^{2}P_{1/2}$ Lamb shift
of Hydrogen.\cite{four}  As formulated by well-known axioms of QFT, 
the vacuum state contains no
on-shell particles; and for QED, in particular, the vacuum 
expectation value $(vev)$ of the
current operator 
$j_{\mu}(x) = ie\cdot \bar{\psi}(x) \gamma_{\mu} \psi (x)\,$ 
 (renormalization and split-point arguments 
                    aside) must vanish in the 
absence of classical, external fields, designated by $A_{\mu}^{ext}(x)$,
\be
< 0 \vert j_{\mu} (x) \vert 0 >_{A^{ext} = 0} = < 0 \vert j_{\mu} (0) \vert 0 >_{A^{ext} = 0 } = 0.
\ee

	When such a classical $A_{\mu}^{ext}$ is present, the current it 
	induces in the vacuum can be 
non-zero,		
$$
< 0 \vert j_{\mu} (x) \vert 0 >_{A^{ext} \not= 0} \not\equiv 0\, ,
$$
although strict current conservation demands that \newline
$\partial_{\mu}
 < 0 \vert j_{\mu} (x) \vert  0 >_{ A_{\not= 0}^{ext}} = 0$     

	The conventional mathematical apparatus used to describe 
	the vacuum state, or vevs of 
current operators, makes no reference to the scales on which vacuum 
properties are to be
observed; and in this sense, it is here suggested that the conventional 
description is
incomplete.  On distance scales larger than the electron's Compton 
wavelength,  $\lambda_e \sim  10^{-10} cm$,
one can imagine that the average separation of virtual $e+$ and $e-$
 currents is not
distinguishable, and hence that the vacuum displays not only a zero
 net charge, but also a
zero charge density.  But on much smaller scales, such as $10^{-20} cm$, 
the average separation
distances between virtual $e+$ and $e-$ are relatively large, with such 
currents describable as
moving independently of each other - until they annihilate.  Since 
there is nothing virtual,
or ``off-shell'', about charge, on sufficiently small space-time scales 
such ``separated"
currents can be imagined to produce effective ``external" fields, 
characterized by an
$A_{\mu}^{ext}(x)$, which could not be expected to be measured at distances
 larger than  $\lambda_e$    	, but
which exist - and contain electromagnetic energy - on scales much 
smaller than  $\lambda_e$.      

	We therefore postulate that, at such small scales and in the 
	absence of conventional, 
large-scale $A_{\mu}^{ext}, < 0\vert j_{\mu}(x)\vert 0 >$
 need be neither $x$-independent nor zero; 
but, rather, that it
generates an $A_{\mu}^{ext}(x)$ discernible at such small scales, 
which is given
 by the conventional
expression
\be A_{\mu}^{ext} (x) = \int d^4 y \, D_{c,\mu\nu}
 (x-y) < 0 \vert j_{\nu} (y) \vert 0 > \, ,
\ee			    
where $D_{c,\mu \nu} $  is the conventional, free-field, causal photon 
propagator that, for 
convenience, is defined in the Lorentz gauge,
$$
0 = \partial_{\mu} A_{\mu}^{ext} = \partial_{\mu} D_{c,\mu\nu} = \partial_{\nu} D_{c,\mu\nu} \, . 
$$

	For comparison, note that classical electromagnetic 
	vector potentials can always be 
written in terms of well-defined, classical currents $J_{\mu}$        ,
 by an analogous relation, ${\cal A}_{\mu} = \int D_{c,\mu \nu} J_{\nu} $, 
while the transition to QED in the absence of conventional, 
large-scale, external fields 
involves the replacement of the classical vector potential
 and currents by operators $A_{\mu}(x)$ 
and $j_{\mu}(x) =  ie\bar{\psi} (x) \gamma_{\mu} 
 \psi (x)$, which satisfy operator equations of motion, 
$(-\partial^2 ) A_{\mu} (x) = j_{\mu} (x)$, 
or			
\be											A_{\mu} = \int D_{c,\mu\nu} j_{\mu} + A_{\mu}^{IN} \, .
\ee

Calculating the vev of (2.3) yields
\be												< 0 \vert A_{\mu} (x) \vert 0 > = \int D_{c,\mu\nu} (x-y) < 0 \vert j_{\nu} (y) \vert 0 > \, ,
\ee
and conventionally, in the absence of the usual, large-scale external 
fields, both 
sides of (2.4) are to vanish.  However, if we assume that non-zero
 $ <0\vert j_{\mu}(x)\vert 0>$  can exist on
ultra-short scales, then a comparison of (2.4) with (2.2) suggests 
that the $A_{\mu}^{ext}(x)$ 
produced by such small-scale currents are to be identified with 
$<0\vert A_{\mu} (x)\vert 0>$ found in
conventional QED in the presence of the same $A_{\mu}^{ext}(x)$.    
In other words, 
\beq
A_{\mu}^{ext} (x) & = & < A_{\mu} (x) > = {1\over i} \int D_{c,\mu\nu} (x-y) {\delta\over \delta A_{\nu} (y)} \cdot e^{{i\over 2} \int {\delta\over \delta A} D_c {\delta\over \delta A}}\cdot \nonumber\\
& & \cdot {e^{L[A+A^{ext}]}\over <S[A^{ext}]> }
\vert_{A\rightarrow 0} \, ,
\eeq				
which provides a bootstrap equation with which to determine such 
short-scale$ A_{\mu}^{ext}(x)$, if
any exist.  In (2.5), which can be transformed into a 
functional-integral relation, the
vacuum-to-vacuum amplitude is given\cite{five} by 
\be
< S[A^{ext}] > = e^{-{i\over 2} \int {\delta\over\delta A} D_c {\delta\over\delta A} } \cdot e^{L[A+A^{ext}]} \vert_{A\rightarrow 0} \, ,
\ee
with $L[A] = Tr \ln [1-ie \gamma\cdot  A S_c  ]$. 

	Of course, one immediate solution to (2.5) 
	is $A_{\mu}^{ext}(x) = 0 = \newline
<A_{\mu}>$, the conventional
solution.  But we are interested in, and shall find, 
solutions that may be safely neglected
at conventional nuclear and atomic distances, but which are 
non-zero in an interesting way
at much smaller distances.  In a sense, such non-zero solutions 
are akin to those found in
symmetry-breaking processes, such as spontaneous or induced 
magnetization; qualitatively
similar ideas have previously been discussed elsewhere,\cite{six} for other 
reasons. We argue
below that these short-scale $A_{\mu}^{ext}(x)$ could be spontaneous, or 
induced by other, virtual,
QCD processes.

 {\bf\section{Approximation}}
 
	How does one go about finding a solution to (2.5)?  
	The first requirement is a
representation for $L[A+A_{\mu}^{ext}]$ which is sufficiently transparent 
to allow the functional
operation of (2.5) to be performed; but, sadly, this is still a 
distant goal, which has been
approximately realized in only a few, special cases.\cite{seven}  What 
shall first be done here is
to use the simplest (perturbative) approximation to L, 
$$
L[B] 
 \Rightarrow {i\over 2} \int B_{\mu} K_{\mu\nu} B_{\nu} \, ,
$$
where the renormalized\cite{five} $K_{\mu\nu}$   corresponds to using only the 
simplest, order-$e^2$, 
closed-electron-loop,
$$
\tilde{K}_{\mu\nu} (k) = \left( \delta_{\mu\nu} - k_{\mu} k_{\nu}/k^2 \right) \cdot k^2 \pi (k^2 ), \quad k^2 = \vec{k}^2 - k_0^2 \, ,
$$
and
\be
\pi (k^2) = {2\alpha\over \pi} \int_0^1 dx \cdot x (1-x) \ln \left( 1 + x (1-x) {k^2\over m^2} \right) \equiv {2\alpha\over\pi} \phi \left( k^2/m^2 \right) \, ,
\ee
where $\alpha  = e^2/4\pi \simeq     1/137 $ denotes the renormalized coupling, 
and $m $ is the electron
mass.  Higher-order perturbative terms should each yield a 
less-important contribution to
the final answer, although the latter could be qualitatively 
changed by their sum; we shall
assume that this is not the case, and that the perturbative 
approximation (properly
unitarized by the functional calculation which automatically 
sums over all such loops) gives
a qualitatively reasonable approximation.

	The functional operation of (2.5), now equivalent to
	 Gaussian functional integration, is
immediate and yields the approximate relation for this 
$A_{\mu}^{ext}(x)$, 
$$
A_{\mu}^{ext} (x) = \int d^4 y \left( D_c K {1\over 1-D_c K}\right)_{\mu\nu} (x-y) \, A_{\nu}^{ext} (y) \, ,
$$
or
\be
\tilde{A}_{\mu}^{ext} (k) = \left( \pi (k^2) \, {1\over 1-\pi (k^2 )}\right) \tilde{A}_{\mu}^{ext} (k) \, .
\ee
A non-zero solution to (8) may be found in the ``tachyonic" form
\be
\tilde{A}_{\mu}^{ext} (k) = C_{\mu} (k) \delta (k^2-M^2 )\, ,
\ee
which then requires that
$$
1 = \pi (M^2 ) \left( 1 - \pi (M^2) \right)^{-1} \, , \pi (M^2) = {1\over 2} \, , 
$$
or $\phi \left( {M^2\over m^2}\right) = {\pi\over 4 \alpha}$, 
which serves to determine $M$.  Note that 
a solution of form $ C_{\mu} \delta (k^2 + \mu^2 ) \,$  would not be possible, 
since the log of $\pi (-\mu^2 )\,$ picks up an imaginary
contribution for time-like $k^2 = - \mu^2$  , for $\mu > 2m$.

	An elementary evaluation of the integral of (3.7)
	 for large $ M/m$ yields
\be
\phi \left( {M^2\over m^2}\right) \simeq \ln \left( 1 + \left({M\over 2m}\right)^2\right) + 0 (1) \simeq 2 \ln \left( {M\over 2m} \right) + \cdots ,
\ee
so that $M  \simeq    2m \exp [\pi/8 \alpha ]$, neglecting relatively small corrections.  
Since $\pi/8 \alpha \simeq        
53.8$, (3.10) yields a value for $M $ close to the Planck mass,  
$ M  \sim     2\times10^{20} GeV/c^2$.  Because
the Physics at distances  $\sim    M_p^{-1}$ - and, indeed, the continuum
 nature of space and time -
is still a matter of speculation, where processes well-defined at 
larger scales are no
longer meaningful, one must exclude intervals smaller than $M_p^{-1}$ when 
calculating any
quantity, such as the electromagnetic energy contained in the fields 
corresponding to the 
$A_{\mu}^{ext}(x)$ of (3.9).

	If the muon and tau self-energy bubbles are included,\cite{eight} the  
	numerical situation changes,
and somewhat more favorably, because the value of $M$ found above decreases 
to $1.2 \times 10^8 GeV$;
this is still large enough to correspond to wave-numbers much larger than 
those associated
with atomic or nuclear distances. However, the upper cut-off on the 
integral of (4.14),
below, is still the Planck mass, $M_p$, and generates the same
order-of-magnitude from which
the possible $d$-values, below, are deduced.  

	The reason that quark-antiquark vacuum-polarization bubbles
	 have not been included is that
those processes would be dominated by strong-coupling gluonic effects, 
which always operate
on time-scales much shorter than those of QED; and hence one might 
expect that $q-\bar{q}$ vacuum
bubbles would be irrelevant to this computation.  It is also quite
 difficult to estimate
such effects, taking into account the appropriate gluonic structure. 
 Of course, one might
argue that at very short distances, the gluonic effects are suppressed
 by asymptotic
freedom, and hence one could employ a $q-\bar{q}$ bubble resembling the leptons',
 except for a
change of charge.  But what does one then use for the quark mass?  
Asymptotic states are not
specified in terms of the quark mass, and a calculation that employs
 $m_q$ should be a
simplifying approximation to much more complicated Physics.  

	In the absence of any firm knowledge, one might choose the 
	smallest values  for $m_{\mu}$ and $m_d$ 
listed in the LBL/CERN Particle Physics Handbook, along with reasonable
 values of the
heavier quarks listed in the same source; and there then follows the 
value $ M\rightarrow 1.6 \times 10^5
GeV$, some three orders-of-magnitude smaller than the $M$ found from all 
the leptons, but still
much larger than atomic or nuclear energies.  For the purposes of this
 note, it does not
seem necessary to include quark bubbles except for the possibility, 
mentioned below, that
rapid QCD fluctuations can serve to provide the initial impetus for 
the QED bootstrap
process. 

{\bf\section{Computation}}

	In order to fully describe this $A_{\mu}^{ext}$, one must define 
	$C_{\mu}(k)$.   The simplest choice would be  
$C_{\mu}\sim   k_{\mu}$; but this corresponds to a ``pure gauge", leading to zero 
 $F_{\mu\nu}$.  If $C_{\mu}\not= 0$, it must
depend on some other 4-vector $v_{\mu}$    , either of origin external to this 
problem, or of origin
within QED, corresponding to an effective phase change.  This question 
will be left open
until the determination of the magnitude of $C_{\mu}$ is made, in comparison 
with the average value
of mass density needed to account for the outward acceleration of the
 universe.

	For the computation of $A_{\mu}^{ext}(x)$, we shall simply choose
	 $C_{\mu}(k) =             
v_{\mu}$, where $v_{\mu}$ is a constant 4-vector (of unknown origin),
 and shall try to justify that
choice below.  Note that $ C_{\mu} \rightarrow v_{\mu} - k_{\mu} (k \cdot v )/k^2 \,$			   would be the more 
appropriate choice in order to retain
the Lorentz gauge condition; but since the additional term is ``pure gauge", 
it does not
contribute to the fields, and will be dropped.  In the familiar 
and convenient ``particle"
units of $c = \hbar =1, \, v_{\mu}$  has the dimensions of length, or inverse mass. 
It will subsequently
be useful to state the dimensions of $v_{\mu}$ in a more general way; and 
a simple, dimensional
inspection shows that the correct dimensions of $C_{\mu}$ are (length)$^{3/2}$  
multiplied by
(energy)$^{1/2}$, so that $v_{\mu}$ may be written as 
\be
\epsilon_{\mu} d^{3/2} \left( \mu c^2\right)^{1/2} = \epsilon_{\mu} \, d^{3/2}\, \left( {\mu\over m} \right)^{1/2} \, \left( mc^2\right)^{1/2} \, ,
\ee
where $\epsilon_{\mu}$     is a 4-vector of unit magnitude,
     $\epsilon^2  =  \pm 1$.  If $c=1$, and $\mu = m$, then an average
value of $d$ (calculated from an average value of the 
magnitude of $v$) is $\sim    10^{-5} cm$; but if $\mu
= M_p$, then the same, average magnitude of $v$ means that 
$d$ is much smaller.  The magnitude of
$v$  will be determined by matching the average energy density
 of the electromagnetic fields
obtained from $A_{\mu}^{ext}(x)$ to the experimental $U $ determined from 
astrophysical data.\cite{two}  These
two extreme values of $d$ may be thought of as defining the scale 
at which the vacuum fields
are generated, or the spatial size at which they can be maintained.  
If these fields of very
high wave-number are spontaneously created, $ d$ can turn out to be 
on the order of $10^{-5} cm$,  
while if these fields are induced by virtual QCD processes, $d$ can 
be on the order of $10^{-13}
cm$.  The present calculation cannot distinguish between these 
possibilities, since any
solution of (3.8) that is multiplied by an arbitrary constant 
is still a solution.

	Of course, the simple replacement $C_{\mu}\rightarrow v_{\mu}$  
	 is insufficient 
	to provide the physical cut off
at frequencies $\sim$     $M_p$; but this could easily be remedied by adjoining 
a factor  $\Theta (M_p - k_0)$ to $v_{\mu}$  in (3.9).  
What shall be done instead is to follow that simplest 
procedure which uses $C_{\mu}= v_{\mu}$ in order to obtain a qualitative 
understanding of $A_{\mu}^{ext}(x)$, and 
its corresponding $E$ and
$B$ fields; but, then, when the total energy of those fields is calculated, 
to introduce the
corresponding cut-off, $M_p$, in the integral (over all space at a given time)
 of the total
electromagnetic energy,	

	Integration of the Fourier transform of (3.9), with $C_{\mu}\rightarrow v_{\mu}$, is straightforward, and yields the ``tachyonic" forms,
\beq
A_{\mu}^{ext} (x) & = & - {2\pi^2 M\over \sqrt{x^2}} \, v_{\mu} N_1 \left( M\sqrt{x^2} \right) \, , \, x^2 > 0 \nonumber\\
& = & - {4\pi M\over\sqrt{-x^2}} \, v_{\mu} K_1 \left( M \sqrt{-x^2}\right) \, , \, x^2 < 0 \\
& = & 0 , \,  x^2 = 0 \, , \nonumber
\eeq
where $x^2 = \vec{x}^2 - x_0^2 \,$,   with regions of propagation and damping, inside 
and outside the light cone
(l.c.), reversed in comparison to those of a causal propagator. 
 [The third entry of (4.11)
represents the Principal Value limit of  $1/x^2$ , as $ x^2\rightarrow 0$.]  
As the l.c. is
approached, from inside or outside, $A_{\mu}(x)$ becomes quite large at intervals 
     $\sim M_p^{-1}$  from
the l.c.; but, as previously noted, quantum gravity - or what may be the
 same thing, a lack
of continuous space-time variables - sets a practical limit to all 
processes at such high
frequencies.  Incidentally, outside the l.c. one makes a distinction 
between frequencies,
 $k_0$, and wave-numbers, $k$,   because $\delta (k^2 - M^2 ) = \delta \left( \vec{k}^2 - [ M^2 + k_0^2 ] \right)$,  
so that $\vert\vec{k}\vert$ is at least as large as $M$, while $k_0$ 
should not exceed $M_p$.

	Inspection of (2.2), with a conserved $<0\vert j_{\mu}(x)\vert 0>$, 
	suggests that one can instead write
$$
A_{\mu}^{ext} (x) = \int D_c (x-y) < 0 \vert j_{\mu} (y) \vert 0 > \, ,
$$
and for the solutions of (4.11), it then follows that
$$\left( - \partial^2 \right) A_{\mu}^{ext} (x) = M^2 A_{\mu}^{ext} (x) = < 0 \vert j_{\mu} (x) \vert 0 > \, .
$$
Hence this $<j_{\mu}(x)>$, as a function of any coordinate system used to describe it, has
propagation only outside the l.c. of that system. A most unusual current! 
 Real currents of
real particles propagate inside l.c.s.

	When the $A_{\mu}^{ext}(x)$ of (4.11) is used to calculate the corresponding 
	$E$ and $B$  fields, and the latter are inserted into 
	$ U = {1\over 8 \pi} (E^2 + B^2)$,   one finds unavoidable 
singularities as the l.c. is
approached; these singularities are effectively removed by refusing to 
consider space-time
intervals $< M_p^{-1}$.  The computation is simpler in momentum space, if we 
calculate an average
energy density defined as the total vacuum electromagnetic energy divided 
by the total
volume of our universe,
$$
\bar{U} = {1\over 8 \pi} \int d^3 x \left( E^2 + B^2 \right) / {4\pi\over 3} R^3 \, ,
$$
and if we simplify matters by integrating $\int d^3x$ over all space, rather 
than fixed by $\int d^3 x =4\pi R^3/3$.  One begins with the Fourier
 representation of (3.9), 
with $C_{\mu} \rightarrow v_{\mu}$; and one then
calculates 
$$
W = \int d^3 x \cdot U
$$
over all space at a fixed time.  This computation requires the 
specification of $v_0^2$      and
the   $v_i^2$   , which enter on the same, multiplicative footing; and 
to make the calculation even
simpler, we imagine that $B = 0, E \not= 0$, so that only electric 
fields are present.  One then
finds
\be
W \rightarrow 4\pi^3 v_0^2 \int_0^{\infty} \, {dk_0\over k_0}\, \left( k_0^2 + M^2 \right)^{3/2} \cdot \cos^2 \left( k_0 x_0 \right) \, ,
\ee
and for $x_0 >> M^{-1}$, replace the $\cos^2(k_0 x_0) $ factor by 1/2, and insert an
 upper limit to the
integral of (4.13), replacing it by
\be
W\rightarrow 2\pi^3 v_0^2 \int_0^{M_{p}} \, {dk_0\over k_0}\, \left(K_0^2 + M^2 \right)^{3/2}\, .
\ee
>From (4.14) it is clear that the dominant effect in any Lorentz 
frame comes from the region
outside of, but on the order of a distance $1/M_p$ from the l.c.; 
and this generates a result
for $W$ proportional to $(M_p)^3$. It is amusing to note that this 
is one power of cut-off less
than would be needed in a computation which sums over all 
permitted, zero-point energies -
but, of course, the Physics here is quite different.  One finds
$$
W \sim {2\over 3} \pi^3 v_0^2 M_p^3 + \cdots
$$
which when divided by 
$$
{4\pi\over 3} R^3 \sim 4 \times 10^{85} cm^3
$$
produces a crude, average value for the total energy density
 (or mass density, when divided
by $c^2$).  When this ratio is equated to the average mass density 
needed to produce the
observed positive acceleration of the universe, 
$U \sim  10^{-29} gm/cm^3$, one finds the value
of $\sim   10^8$ for the dimensionless quantity  $v_0 m$  
(where $m$ is the mass of the electron).  

	From this numerical value follow the values quoted 
	for $d$ in the discussion following
(4.11) From the relation 
$$
v_0 \sim d^{3/2} \mu^{1/2} \, ,
$$
the range of $\mu$ values, from $ m$ to $M $ to $M_p$, define different, 
possible values of $d$.  Table 1
relates these approximate, average quantities.

\begin{center}
\vskip .10in
\begin{tabular}{ |  l | l |}\hline
 $\mu $ &  $d(cm)$  \\ \hline
 $m $ &  $10^{-5}$  \\  \hline
$M $ &  $10^{-8}$  \\ \hline
 $M_P$ & $10^{-13}$ \\ \hline
\end{tabular}
\vskip .10in

Table 1:  Approximate orders of magnitude of $d$ 
for different choices of $\mu$
\end{center}
\vskip .10in

	If $\mu\sim  m$, then the picture that appears is one of weak 
	fields extending over distances on
the order of $10^{-5} cm$; and one might think of these fields as 
occurring spontaneously, in the
sense that there is no outside influence that initially drives
 them.  At the other extreme,
if $\mu\sim  M_p$, then $d \sim  10^{-13} cm$, a typical, strong-interaction 
distance.  It is then tempting to
interpret  $d$  as the scale at which such $A_{\mu}^{ext}(x)$ are generated, 
by a physical, albeit
virtual, process that is external to QED.  Here, the obvious 
candidate is QCD; and one can
imagine, as part of the QCD vacuum structure, virtual quark-antiquark
 pairs, electrically
charged, forming and reforming, with a partially-formed gluonic flux 
tube between them. This
QCD-generated separation of charge, on the order of $10^{-13} cm$, 
effectively ``polarizes the QED
vacuum", and in turn produces a non-zero $A_{\mu}^{ext}$, which the rules of 
QED require to satisfy (3.8).

	Other models, lying in between these two extreme cases are 
	surely possible, but can only be
determined by a more-accurate calculation, one which invokes either 
QED vacuum structure
more complicated than the simplest, closed-fermion-loops of this paper, 
or one which
simultaneously incorporates enough QCD so that the latter can produce
 the $v_{\mu}$ ``spark" that
starts the entire process.  

	The question has  been raised\cite{eight} as to how much significance 
	can be attached to the
replacement of $C_{\mu}(k)$ by a constant $v_\mu$; and the response here is 
somewhat more involved, and
ties in with the form of the calculation performed. The choice of 
constant  $v_{\mu}$ was originally
made for simplicity, for it permits a simple evaluation of all relevant 
integrals.  But if $
d  \sim  10^{-5} cm$ corresponds to spontaneously-appearing fields in a volume 
of that dimension
cubed, then there should exist other ``domains" with other values of $v$. 
However, if the
direction of the spatial v-vector is random - and one would expect this 
if the process is
spontaneous - then an average over the v-directions can produce the same, 
final effect,
since the energies of each region will just add.  Since there is no 
reason to expect
different regions of space-time (that are matter-free) to be different, 
one would also expect that the magnitudes of $v_0$ and $\, \vec{v}\,$ will be unchanged; and this
``justifies" the 
constant-$v$ calculation.

	If, however, $d \sim    10^{-13} cm$, and the entire effect is driven 
	by QCD polarization of the QED
vacuum, then the picture becomes somewhat more reasonable.  Now 
the appearance of a
``constant" $v_{\mu}$ providing the 4-vector index for $A_{\mu}$ no longer raises 
the question of $LI$
violation,  since $v_{\mu}$ is itself ``dynamical", defined by the average 
separation of virtual $q-\bar{q}$ 
pairs; rather, the question now becomes one of a proper characterization 
of the QCD vacuum. 

	What emerges from this ``tachyonic" solution is the intruiging 
	idea that vacuum dynamics is
somehow connected with events that are outside the l.c., in contrast 
to our conventional,
causal world, whose macroscopic dynamics are always on or inside the l.c. 
It should be
emphasized, however, that even if conventional $LI$ is rigorously maintained
 when details of
the QCD vacuum become known, conventional translational invariance of the
 vacuum is open to
question (and redefinition).

	Another question has been raised\cite{nine} concerning the choice of 
	origin of the coordinate
system used.  If observer A uses a coordinate system with spatial origin 
at the tip of the
Eiffel Tower, while observer B uses a coordinate system with spatial 
origin fixed at the
center of Brown University, then one expects that any $ U$ calculated at 
an intermediate point,
at the same time, will display different values to $A$ and $B$, because 
their separation 
distance $ D \sim    5000 km$.  What one might do is to replace $x$
 by $x-x_0$,  
treat $v =v(x_0$), and
average over all $x_0$. But this raises a deeper question, which deserves 
a more thorough
analysis.
	In conventional definitions of an ``external" field it is always, 
	and naturally, assumed
that an external field automatically signifies a lack of translational 
invariance, This is
almost but not precisely the intention of the present case, and for a clear, 
physical
reason: Every observer, regardless of the choice of his space-time 
coordinate system, should
have the same characterization of virtual vacuum structure, and to 
each observer the
$A_{\mu}^{ext}(x)$ measured in terms of his coordinates should be the same.  As 
it happens, to within
an overall, multiplicative 4-vector $v_{\mu}$ - of physical origin external 
to this computation -
this ``effective external" vector potential will be a function 
of the Lorentz-invariant $x^2$,
so that all observers connected by a Lorentz transformation would 
``see" the same solution. 
But something more is intended, if this $A_{\mu}^{ext}$ is to represent an 
unremovable ``external"
field that is seen by every observer because of the QED vacuum 
fluctuations that define his
vacuum. As noted above, this can be arranged mathematically by 
replacing $A_{\mu}^{ext}(x)$ by
$A_{\mu}^{ext}(x-y)$, and averaging over all possible observers' space-time 
points $y$, so that the
desired solution is the same for all observers in terms of the
 $x$-coordinates used by each. 
Such an average is as much a solution to (2.5) as is the $A_{\mu}^{ext}(x)$  
derived and discussed
above.  

	However, if each vacuum state associated with the space-time 
	coordinates of a particular
observer is different - that is, distinguishable - from the vacuum state 
of other observers
at different space-time points, but if they all contain the same physical 
content, and their
forms and predictions are the same for every observer, then an alternate 
and perhaps more
satisfying arrangement is to employ a variant of a unitary transformation (UT) 
to connect
observers at different space-time points. Beginning with an ``absolute" 
vacuum state, $\vert O >_{IN}$,
which is assumed to be the same for all observers, we have written 
$_{IN}< O\vert A_{\mu}(x)\vert O >_{IN} =
A_{\mu}^{ext}(x)$.  However, the external field measured by an observer whose 
coordinate origin is
located at the space-time point $y_{\mu}(i)$ (with respect to an ``absolute" 
coordinate system of
origin, e.g., located at the center of our expanding universe) and 
$x_{\mu}$ as the observation
point (with respect to that ``absolute" origin), would be written as
\beq 
& & <_{IN} 0 \vert A_{\mu} (x-y^{(i)}) \vert 0 >_{IN} = < 0 \vert e^{-iP \cdot y(i)} \, A_{\mu}(x) e^{iP\cdot y^{(i)}} \vert 0 >_{IN} \nonumber\\
& & = < 0; y^{(i)} \vert A_{\mu} (x) \vert A_{\mu} (x) \vert 0 ; y^{(i)} > \, ,
\nonumber
\eeq
where $\vert 0; y > = e^{iP\cdot y} \vert 0 >_{IN}$ and $P_{\mu}$ is the 
complete 4-momentum of 
the entire system, including contributions
from the vacuum fields.  Then, the field measured by an observer 
whose coordinate origin is
located at the (``absolute") coordinate $y_{\mu}^{(j)}$ is simply given by
$$
e^{(y^{(i)} - y^{(j)})\mu {\partial\over\partial x_{\mu}}}  <_{IN} 0 \vert A_{\mu} (x-y^{(i)} ) \vert 0 >_{IN} = <_{IN} 0 \vert A_{\mu} (x-y^{(i)} ) \vert  0 >_{IN} \, ,
$$
and will be different from that measured by the $y_{\mu}^{(i)}$ observer 
if $y_{\mu}^{(i)} \not= y_{\mu}^{(j)}$. 

	To move from frame to frame, one can use this formulation;
	 but within each such frame the
rules of conventional QED apply, and the coordinate difference $(x-y^{(i)}$),
 to the observer in
the frame with origin at $y^{(i)}$, may just be written as: $x$.  In this way, all 
observers are
equivalent, and each measures the same vacuum Physics. Observers using 
different $y^{(i)}$ 
origins will disagree about the value of the vacuum $ A^{ext}$ fields at the 
same (``absolute")
space-time point $x_{\mu}$; but each, when they perform the same physical 
measurement or
computation - such as an estimate of the vacuum fields' total  
energy - will obtain the
same, numerical result.

{\bf\section{Detection}}

	Let us assume that the electric fields following from 
	the $A_{\mu}^{ext}$ of (3.9) are fairly decent
representations of what occurs physically, and consider the 
situation near the l.c., when $E^2 >> m^4/\alpha$    ,
and significant pair production can take place.  
The most relevant, starting
formula is then that of Schwinger's 1951 paper\cite{five} for the 
vacuum persistence probability $P_0$,
\be
\ln P_0 = - {Vt\alpha E^2\over \pi^2} \, \sum_{n=1}^{\infty} \, n^{-2}
 e^{-n\pi m^2/eE}\,,
\ee
where $E$ is a constant electric field across a volume $V$ which has 
existed for a time $t$. 
Analogous forms can be derived\cite{seven} for certain, non-constant fields. 
In the limit of very
large $\vert \vec{ E} \vert = E$, (5.15) can be trivially evaluated as 
$$
\ln P_0 = \Gamma t, \Gamma = {VeEm^2\over 4\pi^2} \cdot \kappa , \kappa = \int_1^{\infty} {dx\over x^2} \, e^{-x} \sim 0 (1) \, ,
$$
so that the vacuum state would disappear exponentially fast into a 
state of many lepton
pairs.  

	What such a computation misses is the effect of the 
	back-reaction, or oppositely-induced
field generated by the emitted pairs, which tends to reduce the 
original $E$.\cite{ten} In the
present case, and even though the fields are not constant, near 
the l.c. one would expect
copious lepton production, leading to a sort of averaged decrease 
of $E$, and an effective plasma or ``thermalization" 
of the combination of effective $E$ and number $n$ of lepton pairs.  
(The previous computation of electrical energy density ``stored" in 
the vacuum is essentially
unchanged, except that that energy is now partitioned between 
weaker fields and the lepton
plasma.)  However, as the l.c. is approached, both the effective $E$ and 
the number n of pairs
must tend to zero, because of the ``experimental" proof that a photon 
{\bf on }the l.c. displays no
interaction with either an effective $E$ nor with the lepton plasma. 

	Could the existence of such a plasma and associated fields 
	near the l.c. be tested? 
Consider the simple example of charged-particle scattering by the 
exchange of a single,
virtual photon.  The amplitude for this process is proportional to 
$(q^2 - i\epsilon  )^{-1}$, and in the
center of mass of the scattering particles, $q_0 = 0$.  Our solution 
for $A_{\mu}^{ext}$ becomes large
near the l.c. when $M^2 x^2 < 1$, and its Fourier transform is proportional to  	
$\delta (\vec{q}^2 - M^2 )$ for $q_0 = 0$.  
In other words, $\vec{E}(q)$ is peaked when $\vert \vec{q}\vert  = M$, which is where the 
plasma region begins,
and where absorption or rescattering by the plasma of this or any 
other virtual photon would
occur.  Hence the presence of such a plasma could be tested - if one 
could ever reach
momentum transfers on the order of $M$ - by observing an anomalous drop 
in the differential
cross section when $\vert \vec{ q} \vert $ exceeds $M$.

	This observation raises the question of the effect of such 
	non-perturbative vacuum
structure on the typical $UV$ divergences of QED.  If all virtual quanta 
at large momenta can
be partially absorbed or rescattered by these vacuum fields, or by the 
plasma resulting from
their essentially non-perturbative pair-production, one can expect an 
effective damping, or
built-in convergence factor, for all sufficiently large momenta $M < q < M_p$.  
In other words, 
a redefined, non-perturbative vacuum could provide the necessary 
convergence factors to
insure finite QFT perturbation expansions, with cut-off parameter $\Lambda$ on 
the order of, or
somewhat larger than $M$. The essential ideas are that such vacuum fields 
exist in any and
every Lorentz frame; and that they become significantly large, 
large enough to produce
pairs, as the l.c. is approached.

{\bf\section{Tachyonic Speculations}}

	We now turn to an even more speculative subject. Let us suppose 
	that vacuum fields at least
qualitatively similar to those suggested above exist, and become large 
as the l.c. is
approached,  both inside {\bf and} outside the l.c. It is therefore natural to 
ask if lepton-pair
production can take place both inside and outside the l.c.  When tearing 
virtual pairs from
the vacuum, what is to prevent some of them from being formed with more 
momentum than
energy?  All of our previous experience in the world in which material 
particles inside the
l.c. can never be driven outside, $v < c$, does not provide an unambiguous
 answer to this
question.  In this Section we ask if such tachyonic creation might be 
possible, and consider
the qualitative results.

	It may be useful to place this discussion in the form of 
	several Questions, which appear in
a natural sequence, and are followed by  tentative Answers.

a) Can lepton pairs be produced outside the l.c. by the electromagnetic 
fields of this 
$A_{\mu}^{ext}(x)$?

b) If so, can one expect such tachyonic leptons to remain tachyonic, 
that is, to remain
outside the l.c. if their subsequent (classical) motion is governed 
by the fields of this 
$A_{\mu}^{ext}(x)$?

c) Could such tachyonic leptons contribute to the ``missing matter" 
of current astrophysical
and cosmological interest?

d) Are there any other events, or signals, that could be attributed 
to such tachyonic matter?

	Initial, but {\bf very} rough estimates, suggest a positive answer 
	to each of these questions, as
follows.

a) A first-estimation of lepton-pair production close to the l.c. by 
this $A_{\mu}^{ext}(x)$ yields a
Schwinger-like $\Gamma$ of (5), in which $\vert E \vert$ is replaced by 
$\vert (x\cdot v)/ (x^2)2\vert$,  
which is the same for
production inside and outside the l.c. In fact, one might expect more 
production outside the
l.c., because these fields fall away exponentially (with increasing argument)
 inside the l.c.

b)  There appear to be solutions to the classical equations of motion
 of a charged tachyon
in this $A_{\mu}^{ext}(x)$, in which the tachyon remains outside the l.c., that is, 
the l.c. relevant to the description of its motion.

c) There is no obvious reason why such tachyonic matter could not 
gradually accumulate and
become a significant fraction of the ``dark" or "missing" matter.  
If their interactions with
ordinary matter and each other are governed by energy and momentum 
conservation, and if
following their production, tachyonic leptons have momenta far greater 
than ordinary
leptons, they would be expected to be spatially well-separated from the 
latter. Note also
that their motion in the magnetic field of $A_{\mu}^{ext}(x) $ suggests that tachyonic
 leptons could be
captured in ``circular orbits" near the outer reaches of the universe, 
forming ``shells" at
galactic distances, and would have a very different type of existence 
from that of ordinary
leptons.

d)  Gamma-ray bursts and their associated $x$-ray residues could be a 
consequence of
occasional, tachyonic lepton-pair annihilation, with oppositely-charged 
particles moving
about their ``shell" in opposite directions, and occasionally colliding; 
they could convert
in their Center of Mass to an extremely high-energy, virtual photon or 
conventional pair
that subsequentially annihilate with each other, or with other, conventional 
matter.  This could also be a mechanism to produce ultra-high-energy cosmic rays.  

	Of course, these are all speculations, especially c) and d);
	 but the existence of such
$A_{\mu}^{ext}(x)$  does suggest possibilities which do not seem to have been explored. 
 One emphasizes
that the arguments which have been used to obtain the positive 
Answers presented above are
crude; and better calculations may well reverse those conclusions.  But
 it does seem as if
these are Questions that are worth asking.

{\bf\section{Summary}}
	
	There are other, less-dramatic and numerically less-important 
	solutions to (3.8), which
might be mentioned.  For example, the choice
$$
C_{\mu} (k) = \xi \, v_{\mu} \delta (k\cdot v)
$$
rigorously maintains the Lorentz gauge condition; and, with $v_{\mu}$ timelike, 
has no singularity
on the l.c., and is non-zero only outside the l.c., 
$$
A_{\mu}^{ext} (x) = \xi {v_{\mu}\over\sqrt{-v^2}} \, M \phi (MX) \, , \, \phi (u) = {\sin u\over u} \, ,
$$
where $X = (x^2 - (x\cdot v)^2)^{1/2}$ and $\xi$    
is a constant.  Rotating $v_{\mu}$ to 
lie along the 4-direction
produces the simple, time-independent result, 
$A_{\mu}(x)\rightarrow  i A_o     ,  A_0 =  \xi   (2\pi/r) 
\sin(Mr)$ , 
with $r =\vert \vec{x}\vert$.  In order to have fields from 
such a solution generate 
the needed acceleration
of the universe,  $\xi$     would have to be chosen extremely large.

	Whichever the form of appropriate $A_{\mu}^{ext}$ chosen, 
	the essential 
	and new physical idea
proposed above is that on sufficiently small space-time scales, 
induced vacuum currents
should generate  effective, external fields, fields that are associated
 with vacuum
structure, and have measurable consequences.  It is difficult not to  
imagine that these
electromagnetic fields, residing principally outside any l.c., are the 
source of the
astrophsicists' missing, ``dark energy"; while the lepton-pairs, produced 
close to the l.c.,
inside or outside the l.c., can contribute to the missing ``dark" matter.  
It should be noted
that if the highly-nonlinear equations of motion of a newly-produced
 charged particle (in an
``external" field that is extremely large near the l.c.) turn out to be 
chaotic, one would
have a possible mechanism for the diffusion of these leptons throughout 
the universe.

	One cannot deny that the calculations and estimates above are crude.
	 Nevertheless, this
mechanism for generating a significant vacuum energy from vacuum-induced
 electromagnetic
fields of very high wave numbers and frequencies may have a certain reality. 
 One should
realize that the energy produced by such virtual currents is real; 
and that crudeness does
not necessarily preclude correctness.  
\vskip .10in

\noindent{\bf NOTES}

	This paper is based upon a previous, unpublished contribution 
	(arXiv:hep-th/0303108) in
which this general bootstrap idea was first put forward.  Additional, 
clarifying material
has been added, along with the tachyonic speculations; and some of the 
numerical estimates
presented there have been sharpened.


\begin{thebibliography}{**}
\bibitem{two} The author is indebted to I. Dell'Antonio for this numerical 
estimate, as well as for several illuminating conversations.
\bibitem{nine}  The author is indebted to H. de Vega for this, as well 
as other, most-constructive
criticism.
\end{thebibliography}
\end{document}